# Improving the Representation of Energy Efficiency in an Energy System Optimization Model

Neha Patankar[1], Harrison G. Fell[2], Anderson Rodrigo de Queiroz[3,4], John Curtis[5], Joseph F. DeCarolis[3,*]

[1]Andlinger Center for Energy and the Environment, Princeton, US

[2]Department of Agricultural and Resource Economics, NC State, Raleigh, US

[3]Department of Civil, Construction and Environmental Engineering, NC State, Raleigh, US

[4]Department of Decision Sciences, School of Business at NC Central University, Durham, US

[5]Economic and Social Research Institute, Dublin, Ireland

*Corresponding author. Phone: +1 919-515-0480; Fax: +1 919-515-7908; Email: jdecarolis@ncsu.edu

**Abstract**

Energy system optimization models (ESOMs) are designed to examine the potential effects of a proposed policy, but often represent energy-efficient technologies and policies in an overly simplified way. Most ESOMs include different end-use technologies with varying efficiencies and select technologies for deployment based solely on least-cost optimization, which drastically oversimplifies consumer decision-making. In this paper, we change the structure of an existing ESOM to model energy efficiency in way that is consistent with microeconomic theory. The resulting model considers the effectiveness of energy-efficient technologies in meeting energy service demands, and their potential to substitute electricity usage by conventional technologies. To test the revised model, we develop a simple hypothetical case and use it to analyze the welfare gain from an energy efficiency subsidy versus a carbon tax policy. In the simple test case, the maximum recovered welfare from an efficiency subsidy is less than 50% of the first-best carbon tax policy.

1. Introduction

In order to avert the worst effects of climate change, the IPCC indicates that the world needs to achieve net-zero carbon emissions around the middle of this century (IPCC, 2018). While significant policy effort and supporting analysis has been focused on supply-side clean energy, demand-side energy efficiency also represents a critical mechanism to reduce energy and greenhouse emissions (Creutzig et al., 2018). Over the last few decades, strong efficiency gains have produced a significant impact on global energy demand, reducing consumer energy bills (Sorrell, 2015), holding back emissions growth (IEA, 2021), and making energy systems more secure by reducing the dependency on energy imports (Gillingham et al., 2009). Advocates of ambitious climate policies often support simultaneously imposing a price on carbon and alternative policies, such as renewable portfolio standards (RPS), which sometimes credit energy efficiency (Baranzini et al., 2015). For example, energy efficiency crediting was one of the means to comply with the intensity standards under the US Environmental Protection Agency's (EPA) Clean Power Plan (CPP). Assessing the efficacy of such policy is challenging: it is not always clear how policies that include energy efficiency crediting or subsidies compare to the first-best solution under a Pigouvian tax.

To address this issue, Fell et al. (2017) develop a novel model that considers the tradeoff between expenditures on energy efficiency versus electricity supply in a manner that is theoretically consistent with microeconomic theory. In their formulation, both households and firms explicitly consider both energy consumption and energy efficiency to meet service demands, where investments in energy efficiency are treated as avoided consumption. The authors find that optimally crediting energy efficiency under an emissions intensity standard can recover the first-best outcome under an assumption of inelastic service demands, but not when those service demands are assumed to be elastic. More broadly, their model formulation can assess the welfare implications of various policy measures that incorporate energy efficiency. The goal of this study is to expand and incorporate their formulation into an energy system optimization model (ESOM), which employs linear optimization to perform capacity expansion across an energy system in order to develop projections of technology deployment, emissions, and cost. Further details on the formulation of ESOMs are given in Section 2. Incorporating the formulation by Fell et al. (2017) into an ESOM provides the ability to examine the welfare effects of energy efficiency measures along with other policy alternatives over time. This work represents a critical methodological advancement since ESOMs are a key tool used to evaluate deep decarbonization pathways that ultimately inform policy.

In previous work, top-down and bottom-up modeling approaches have been used to model the system-wide effects of energy efficiency (Van Beeck et al., 2000), and their contrasting styles have led to divergent projections of technological change and the cost of that change (Horne et al., 2005). Top-down approaches typically take an aggregate view of the economy and consider market distortions, income effects, and the relation between various economic agents, such as households and government. By contrast, bottom-up approaches using ESOMs represent individual technologies so that changes in the technology mix can be modeled explicitly. ESOMs typically minimize the present cost of energy supply by deploying and utilizing energy technologies over time to meet a set of exogenously specified end-use demands. Although rich in technology detail, exogenously specified end-use demands restrict the feedback effect on the consumer side. Exposing electricity end-users to varying prices inevitably results in behaviors that maximize consumer welfare (Nardelli et al., 2017).

Several features have been incorporated into ESOMs to better represent demand response and end-use energy efficiency. First, many ESOMs include piece-wise linear demand curves, which map the quantity demanded of energy services to their market price, rather than exogenously specified service demands (Kirchem et al., 2020). Second, the model input datasets often include a suite of different technologies with varying levels of energy efficiency. The ESOM then selects efficient technologies and the level of demand response based on the specified price elasticities of service demand (Božić, 2007). Third, assuming exogenous efficiency ratios higher than the baseline can be used to represent a higher penetration of efficient technologies (Yanbing and Qingpeng, 2005). Fourth, expert knowledge can be used to assume exogenous technology adoption targets driven by energy efficiency policy (McNeil, 2008). These approaches can often lead to prescriptive results that are unrealistic. For example, least cost optimization selects only the most efficient technologies (e.g., LED lighting and subcompact cars). Modelers often then add hurdle rates to control the rate of efficient technology adoption; however, there is a little empirical basis for the choice of hurdle rate values (DeCarolis et al., 2017). There are some recent efforts to model market heterogeneity, consumer behavior, and intangible costs. For example, van Zoest et al. (2021) quantify the response of different consumer types to a compulsory demand charge in the Swedish commercial sector, and Diao et al. (2016) model the intangible costs of traffic policies on electric vehicles in China. Reviewing modeling efforts related to energy efficiency policy, Mundaca et al. (2010) conclude that the modeling and evaluation of policy instruments addressing consumer behavior remains a major challenge for the energy modeling community.

The approach described in this paper represents a significant methodological advancement over previous ESOM modeling efforts aimed at improving the representation of energy efficiency and allows us to systematically evaluate the welfare implications of different policies related to energy efficiency. The formulation for the first time establishes a direct linkage between an energy efficiency subsidy, energy consumption, service demands, as well as consumer and producer welfare in an ESOM. We utilize Tools for Energy Model Optimization and Analysis (Temoa), an open-source ESOM, for this exercise. We compare the welfare gains associated with a carbon tax, representing the first-best policy, versus a subsidy for energy efficiency, which represents a second-best policy. Sensitivity analysis is performed on selected parameters to analyze the effect on the overall welfare gain. The resultant model formulation presents challenging computational issues, as it introduces non-linearities into Temoa's objective function and constraints. We refer to the restructured model as "Temoa-EE+" throughout the paper.

The rest of the paper is organized as follows. Section 2 describes ESOMs and their economic interpretation. Section 3 outlines the Temoa-EE+ mathematical formulation to consider the substitution between electricity and energy efficiency, while Section 4 describes a hypothetical test case used to illustrate the effects of substitution on the modeled system. Section 5 describes how an energy efficiency subsidy can be compared to a carbon tax policy. Section 6 presents our results and discussion, and Section 7 presents our conclusions and outlines future work to apply this enhanced framework.

## 2. Introduction to ESOMs

To conduct the analysis in this paper, we utilize Tools for Energy Model Optimization and Assessment (Temoa), an open-source ESOM. The model formulation is detailed in Hunter et al. (2013), and the Temoa source code is publicly available on Github (Temoa Github, 2021). A snapshot of the code and data used to conduct this analysis is also available through Zenodo, a publicly accessible archive (https://zenodo.org/record/3678734), which allows other researchers to replicate our results and utilize our implementation of the Temoa-EE+ model, as described in the following sections. Table 1 summarizes the nomenclature used for the Temoa-EE+ model formulation.

**Table 1**: Temoa-EE+ model nomenclature

| A. Indices | |
|---|---|
| $t$ | *Index of model time period* |
| $i$ | *Index for technologies* |
| $v$ | *Index of technology vintages* |
| **B. Sets** | |
| $I$ | *Technologies* |
| $V_i$ | *Vintages associated with technology i* |
| $T_i$ | *Time periods associated with technology i* |
| $T$ | *All model time periods* |
| **C. Parameters** | |
| $\alpha$ | *Productivity of energy efficiency in the production of energy services (ranges from 0 - 1)* |
| $\sigma$ | *Elasticity of substitution between electricity and energy efficiency* |
| $\epsilon$ | *Price elasticity of energy service demand* |
| $\gamma_{i,v,t}$ | *Emission activity associated with technology i of vintage v in time period t* |
| $P\theta$ | *Marginal cost of energy efficiency* |
| $E_t^0$ | *Reference electricity demand in time period t* |
| $PE_t^0$ | *Reference electricity price in time period t corresponding to $E_t^0$* |
| $\varphi_t$ | *Constant derived from $E_t^0$ and $PE_t^0$* |
| $ES_t^{min}$ | *Lower bound of energy service demand* |
| $B$ | *Coefficients of all the other ESOM' constraints* |
| $b$ | *Right hand side of all the other ESOM' constraints* |
| $IC_{i,v}$ | *Nominal investment cost associated with technology i of vintage v* |
| $FC_{i,v,t}$ | *Nominal fixed cost associated with technology i of vintage v in time period t* |
| $VC_{i,v,t}$ | *Nominal variable cost associated with technology i of vintage v in time period t* |
| $D_t$ | *Demand in time period t* |
| $\beta$ | *Efficiency credit (ranges from 0 - 1)* |
| $\zeta_{i,v,t}$ | *Factor converting $CAP_{i,v}$ to $ACT_{i,v,t}$* |

| D. Variables | |
|---|---|
| $CAP_{i,v}$ | Capacity associated with technology i of vintage v in time period t |
| $ACT_{i,v,t}$ | Activity associated with technology i of vintage v in time period t |
| $X$ | All other variables in ESOM |
| $E_t$ | Quantity demanded of electricity in time period t |
| $\theta_t$ | Quantity demanded of energy efficiency in time period t |
| $PE_t$ | Electricity price in time period t |
| $P_t$ | Marginal price of energy service demand in time period t |
| $ES_t$ | Energy service demand in time period t |
| E. Functions | |
| $Q_Y$ | Quantity demanded of Y |
| $P_Y$ | Price of Y |
| $f(E,\theta)$ | Dummy function used for describing elasticity of substitution |
| $U$ | Utility function |
| $e$ | Expenditure function |
| $g(ES)$ | Dummy function used for describing utility of energy services |

ESOMs such as Temoa are widely used to analyze energy system capacity expansion plans and employ scenario analysis to investigate different technical, economic, and policy assumptions. The energy system is described algebraically as a network of linked processes that convert raw energy commodities (e.g., coal, oil, biomass) into end-use demands (e.g., lighting, transport, water heating) through a series of one or more intermediate energy forms (e.g., electricity, gasoline, ethanol). Each process is defined by a set of engineering, economic, and environmental characteristics (e.g., capital cost, fixed and variable operations and maintenance cost, efficiency, capacity factor, emission factor) associated with converting an energy commodity from one form to another. Processes are linked together in a network via model constraints representing the allowable flow of energy commodities. The objective of ESOMs is to minimize the present cost of energy supply by utilizing energy processes and commodities over a user-specified time horizon to meet a set of exogenously specified end-use demands. ESOMs simultaneously make technology investment decisions and operating decisions while maintaining an energy balance between primary energy resources, secondary fuels, final energy consumption, and end-use energy services. ESOMs are typically formulated as linear programming models in which technology capacity is utilized to meet end-use demands.

Assuming a single exogenously specified end-use demand, a simplified ESOM with an objective to minimize total system cost can be written as the following linear program:

$$\min \sum_{v \in V_i} \sum_{i \in I} IC_{i,v} \, CAP_{i,v} + \sum_{t \in T_i} \sum_{v \in V_i} \sum_{i \in I} FC_{i,v,t} \, CAP_{i,v} + \sum_{t \in T_i} \sum_{v \in V_i} \sum_{i \in I} VC_{i,v,t} \, ACT_{i,v,t} \quad (1)$$

$$\text{s.t.} \sum_{v \in V_i} \sum_{i \in I} ACT_{i,v,t} \geq D_t \qquad \forall \, t \in T \quad (1a)$$

$$\zeta_{i,v,t} \cdot CAP_{i,v} \geq ACT_{i,v,t} \qquad \forall \, t \in T_i, v \in V_i, i \in I \quad (1b)$$

$$B \cdot X \geq b \quad (1c)$$

In the above formulation, $V, I$ and $T$ are the set of all vintages, technologies and model time periods, respectively, where, $v$, $i$ and $t$ are the indices of these sets. $IC_{i,v}, FC_{i,v,t}$ and $VC_{i,v,t}$ are the discounted investment cost, fixed operations and maintenance cost, and variable operations and maintenance cost of technology $i$, respectively. $CAP_{i,v}$ is the decision variable representing the installed capacity of technology $i$ of vintage $v$. In the above model formulation, the total commodity production from a process is referred to as "activity," $ACT$. Thus, $ACT_{i,v,t}$ is the decision variable representing output of technology $i$ of vintage $v$ in time period $t$. $CAP_{i,v}$ and $ACT_{i,v,t}$ are two inherently different units of measure. $CAP_{i,v}$ represents installed capacity expressed in units of power, while $ACT_{i,v,t}$ represents energy production. Moreover, $\zeta_{i,v,t}$ is a composite factor that converts available capacity to maximum available activity, $\zeta_{i,v,t} \cdot CAP_{i,v}$. Temoa constrains the activity variable $ACT_{i,v,t}$ such that it does not exceed the maximum production possible given $CAP_{i,v}$. $D_t$ is the end-use demand in time period $t$. Furthermore, $B$ represents the coefficients of all the other constraints, and $b$ represents the right-hand side of these constraints. The equations can thus be interpreted as follows: (1) expresses the total discounted system cost to be minimized, (1a) is the set of demand satisfaction constraints, where the right-hand side represents the exogenous demand to satisfy, (1b) denotes the relation between available capacity and activity, and (1c) is the set of all other constraints. Hunter et al. (2013) provide a detailed formulation for the constraints included in (1c). We use this highly simplified algebraic formulation as a starting point and focus on the changes required to model the tradeoff between electricity supply and energy efficiency.

## 2.1. Economic Interpretation of ESOMs

The ESOM formulation, as given in (1)-(1c), meets exogenously specified end-use demands at the minimum system cost. In this formulation, a mix of individual technology outputs produces the required sectoral output (e.g., billion kilometers of heavy truck service or petajoules of residential cooling service). In this paper, we use the concept of welfare maximization, which extends the cost minimization approach used by many ESOMs. We maximize the total consumer and producer surplus over the model time horizon by using a demand elasticity to model a price-responsive demand.

Moreover, we use the concept of a production function, which defines the physical relationship between end-use services and energy commodity inputs to a sector. Production functions are implicitly constructed in cost-minimizing ESOMs based on the optimal selection of technologies to meet demand. For example, electricity production is determined endogenously based on the cost-effectiveness of electricity compared to other fuels and the cost and performance specifications of different generators types. In this paper, we explicitly define a production function that generates energy service from the provision of electricity and energy efficiency. Section 2.3 provides the formulation of a price-responsive demand and production function for energy services.

## 2.2. Demand Elasticities and Elasticities of Substitution

Several ESOMs maximize welfare by including an end-use demand that is responsive to prices. Price responsive demand provides a useful first step in capturing both human behavior and economic feedback to changes in the energy system. These models use demand elasticity to replace exogenously specified demands with inverse demand functions (Loulou and Lavigne, 1996). We extend this effort further by allowing the model to consider the substitution effect between electricity and energy efficiency explicitly. To do so, we provide definitions of demand elasticity and elasticity of substitution in the context of energy system models. The demand elasticity $\epsilon$ of a good, $Y$, is defined as

$$\epsilon_Y = \frac{\frac{dQ_Y}{Q_Y}}{\frac{dP_Y}{P_Y}} \tag{2}$$

where, $Q_Y$ is a quantity demanded and $P_Y$ is the price. From Equation 2, we see that the demand elasticity of $Y$ is the ratio of the percent change in $Q_Y$ to the corresponding percent change in $P_Y$. Measuring the responsiveness of a dependent variable to an independent variable in percentage terms rather than simply as the derivative of the function has the attractive feature that this measure is invariant to the units of the independent and the dependent variables. In this paper, we use price elasticity to specify the responsiveness of demand of energy service to its price.

Now we introduce the elasticity of substitution for a function of two variables. The elasticity of substitution is most often discussed in the context of production functions, which defines the relationship between quantities of input and output goods. The elasticity of substitution considers two-factor inputs to a utility or production function. It measures the percentage response of the relative marginal products of the two factors to a percentage change in the ratio of their quantities. To define the elasticity of substitution, we represent the utility function, $U$, as a function of the quantity demanded of energy services, $ES$:

$$U = g(ES) \tag{3}$$

Since, the quantity demanded of energy services is a function of quantity demanded of electricity, $E$, and energy efficiency, $\theta$, the utility, $U$, can be given as $g(f(E,\theta))$. Then the elasticity of substitution between electricity and energy efficiency is given by:

$$\sigma_{\theta E} = -\frac{d(\theta/E)}{\theta/E} \Bigg/ \frac{d\left(\frac{dg}{d\theta}\Big/\frac{dg}{dE}\right)}{\frac{dg}{d\theta}\Big/\frac{dg}{dE}} \tag{4}$$

A special class of production functions includes a constant elasticity of substitution (CES), $\sigma$. CES production functions were first explored by Arrow et al. (1961), who proved that a production function with two inputs has a constant elasticity of substitution $\sigma$ between inputs if and only if the production function is either of the functional form:

$$f(E,\theta) = (\alpha \cdot \theta^\rho + (1-\alpha) \cdot E^\rho)^{1/\rho} \tag{5}$$

or else of the Cobb-Douglas form, when elasticity of substitution is unity:

$$f(E,\theta) = (\theta^\alpha \cdot E^{1-\alpha}) \tag{6}$$

The parameter $\alpha$ represents share of an input, $0 \leq \alpha \leq 1$ and $\rho$ is a constant equal to $(\sigma - 1)/\sigma$. Electricity supply and energy efficiency are substitute goods, i.e., one good can be used in place of other. As a result, the elasticity of substitution between electricity and energy efficiency is greater than one. Thus, we use the production function given by Equation (5) for the formulation in Section 3.

## 3. Representation Within an ESOM

The conceptual starting point for the restructured model is the flow of energy commodities and money in a simplified economy, as shown in Figure 1. The first group of actors in the diagram is consumers, who pay for energy-efficient technologies and electricity in order to receive energy services. Producers represent the second group of actors. In this case, electric utilities invest in the electricity generation technologies required for the creation of energy services demanded by consumers. In tracing the circular flow, one can start with the utilities, who make investments that supply electricity to consumers. Consumers then pay for both energy efficiency and electricity in order to satisfy their demand for energy services.

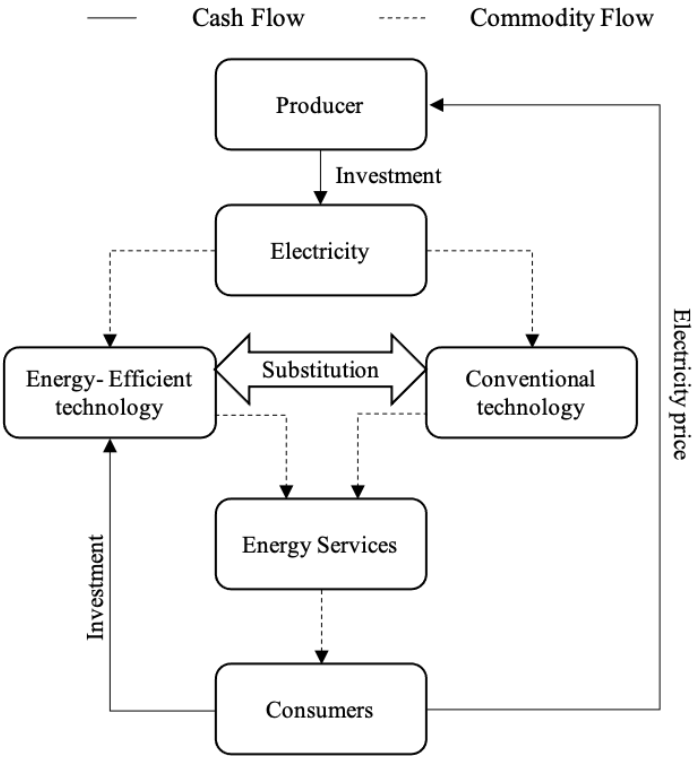

**Figure 1:** Conceptual cash and commodity flows associated with the proposed representation of energy efficiency in a restructured ESOM.

Even though we are not considering other factors in the economy such as labor, wages, and the circulation of earnings, economic equilibrium represented in Figure 1 results in the conservation of both product and value. The difference between payment from consumers and the cost of production for utilities is the producer surplus (profit), while the difference between the consumer's willingness to pay for the energy services and what the consumer actually pays is the consumer surplus. In this way, the model maximizes both producer and consumer surplus. We assume that customers consume a combination of electricity and energy efficiency to maximize their utility, which results in the maximization of consumer surplus. Similarly, the producer maximizes profits, or equivalently, producer surplus, by choosing the appropriate electricity generation and energy efficiency investment. In general, the model maximizes the total welfare, which is the sum of producer surplus and consumer surplus, as shown in Figure 2.

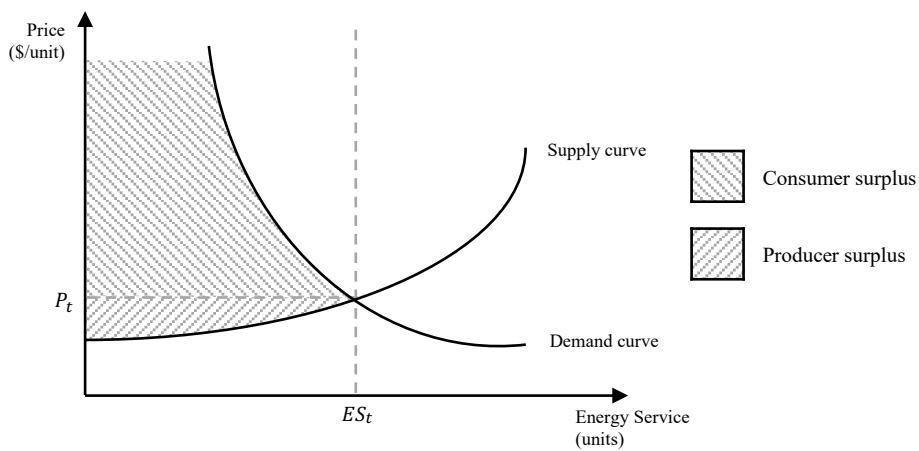

**Figure 2:** Supply-demand equilibrium for energy services. Note that the supply curve of energy services is a function of the supply curves for electricity and energy efficiency. Similarly, the demand curve of energy services is a function of the demand curves for electricity and energy efficiency. The dashed line shows price $P_t$ and quantity $ES_t$ at equilibrium for a given time period.

In the model, the consumers' demand for energy services is dependent on the price and quantity demanded of energy efficiency and electricity, which in turn affect one another. This effect is captured by assuming a constant elasticity of substitution production function for the production of energy services.

### 3.1. Temoa-EE+ Model Formulation

In the revised model, the consumption of energy services, $ES_t$, takes two inputs: electricity, $E_t$, and energy efficiency, $\theta_t$. We start with the energy service demand as a function of the energy service price and perform a series of calculations to develop a constraint set consisting of Equations (7), (8), (11), and (12) which are directly implemented in the model. We assume that the quantity demanded of the energy service is inversely proportional to its price, such that the quantity demanded decreases with an increase in the unit price of the energy service. Therefore, we assume that energy service demand, $ES_t$, has a constant own-price elasticity of the form:

$$ES_t = \varphi_t P_t^\epsilon \tag{7}$$

The unit cost corresponding to the energy service demand is given as a function of the electricity price and the energy efficiency price. It is a tedious but straightforward application of calculus to demonstrate that in the CES form (Rutherford, 2002), the unit cost function is given by:

$$P_t = (\alpha^\sigma \cdot P\theta^{1-\sigma} + (1-\alpha)^\sigma \cdot PE_t^{1-\sigma})^{1/(1-\sigma)} \tag{8}$$

As a result, $ES_t$, is given by a convex, differentiable function of the electricity price, $PE_t$, and the energy efficiency price, $P\theta$. Shephard's lemma (Shepherd, 2015) states that demand for a particular good, at a given price, equals the derivative of the expenditure function with respect to the price of the good. The expenditure function, which is the minimum amount spent by consumers on energy services, is hence given as a product of the unit cost of energy services, given in Equation (8), and the quantity of energy services, given by Equation (7):

$$e = ES_t \cdot P_t \tag{9}$$

After substituting Equations (7) and (8) into (9), Equation (9) can be rewritten as:

$$e = \varphi_t \cdot (\alpha^\sigma \cdot P\theta^{1-\sigma} + (1-\alpha)^\sigma \cdot PE_t^{1-\sigma})^{1+\epsilon/(1-\sigma)} \tag{10}$$

Applying Shephard's lemma, we differentiate the expenditure function with respect to $PE_t$ to obtain the electricity demand, and with respect to $P\theta$ to obtain the energy efficiency demand as a function of the electricity price and the energy efficiency price:

$$\frac{\partial e}{\partial PE_t} = E_t = \varphi_t \cdot (1-\alpha)^\sigma \cdot PE_t^{-\sigma} \cdot (\alpha^\sigma \cdot P\theta^{1-\sigma} + (1-\alpha)^\sigma \cdot PE_t^{1-\sigma})^{(\epsilon+\sigma)/(1-\sigma)} \quad (11)$$

$$\frac{\partial e}{\partial P\theta} = \theta_t = \varphi_t \cdot \alpha^\sigma \cdot P\theta^{-\sigma} \cdot (\alpha^\sigma \cdot P\theta^{1-\sigma} + (1-\alpha)^\sigma \cdot PE_t^{1-\sigma})^{(\epsilon+\sigma)/(1-\sigma)} \quad (12)$$

To derive the constant, $\varphi_t$, we substitute historical values for quantity demanded of electricity, $E_t^0$, and the corresponding price, $PE_t^0$, in Equation (11). As a result, $\varphi_t$ can be given by

$$\varphi_t = \frac{E_t^0}{((1-\alpha)^\sigma \cdot (PE_t^0)^{-\sigma} \cdot (\alpha^\sigma \cdot P\theta^{1-\sigma} + (1-\alpha)^\sigma \cdot (PE_t^0)^{1-\sigma})^{(\epsilon+\sigma)/(1-\sigma)})} \quad \forall\, t \in T \quad (13)$$

We assume that the market is competitive, and the optimization problem is set up as follows.

$$\max \sum_{t \in T} \int_{ES_t^{min}}^{ES_t} P_t(q)\, dq - \sum_{t \in T} P\theta \cdot \theta_t - \sum_{v \in V_i} \sum_{i \in I} IC_{i,v} \cdot CAP_{i,v} - \sum_{t \in T_i} \sum_{v \in V_i} \sum_{i \in I} FC_{i,v,t} \cdot CAP_{i,v} \quad (14)$$

$$- \sum_{t \in T_i} \sum_{v \in V_i} \sum_{i \in I} VC_{i,v,t} \cdot ACT_{i,v,t}$$

s.t. $\sum_{v \in V_i} \sum_{i \in I} ACT_{i,v,t} \geq E_t \qquad \forall\, t \in T \quad (14a)$

$\zeta_{i,v,t} \cdot CAP_{i,v} \geq ACT_{i,v,t} \qquad \forall\, t \in T_i, v \in V_i, i \in I \quad (14b)$

$ES_t = \varphi_t \cdot P_t^\epsilon \qquad \forall\, t \in T \quad (14c)$

$P_t = (\alpha^\sigma \cdot (P\theta)^{1-\sigma} + (1-\alpha)^\sigma \cdot PE_t^{1-\sigma})^{1/(1-\sigma)} \qquad \forall\, t \in T \quad (14d)$

$E_t = \varphi_t \cdot (1-\alpha)^\sigma \cdot PE_t^{-\sigma} (\alpha^\sigma \cdot (P\theta)^{1-\sigma} + (1-\alpha)^\sigma \cdot PE_t^{1-\sigma})^{(\epsilon+\sigma)/(1-\sigma)} \qquad \forall\, t \in T \quad (14e)$

$\theta_t = \varphi_t \cdot \alpha^\sigma \cdot (P\theta)^{-\sigma} \cdot (\alpha^\sigma \cdot (P\theta)^{1-\sigma} + (1-\alpha)^\sigma \cdot PE_t^{1-\sigma})^{(\epsilon+\sigma)/(1-\sigma)} \qquad \forall\, t \in T \quad (14f)$

$B\,X \geq b \quad (14g)$

The objective function given in (14) can be divided into three parts: the area under the energy service demand curve represented by $\sum_{t \in T} \int_{ES_t^{min}}^{ES_t} P_t(q)\, dq$, the area under energy efficiency supply curve represented by $\sum_{t \in T} P\theta \cdot \theta_t$, and the area under the electricity supply curve represented as $\sum_{v \in V_i} \sum_{i \in I} IC_{i,v} \cdot CAP_{i,v} + \sum_{t \in T_i} \sum_{v \in V_i} \sum_{i \in I} FC_{i,v,t} \cdot CAP_{i,v} + \sum_{t \in T_i} \sum_{v \in V_i} \sum_{i \in I} VC_{i,v,t} \cdot ACT_{i,v,t}$. We choose an arbitrary lower bound on the integral, $ES_t^{min}$ such that $ES_t^{min} < ES_t$, in the first part of the objective function to prevent consumer surplus from being unbounded as $ES_t \to 0$. The pictorial representation of the demand curve for energy services can be seen in Figure 2. Note that we do not have a direct representation of a supply curve of energy services in the above model. Since the producer invests in electricity and energy efficiency, the supply curve for energy services is endogenously formed as a function of the supply curve of electricity and energy efficiency.

For the optimal value of independent decision variables $\boldsymbol{PE_t}$, $\boldsymbol{CAP_{i,v}}$ and $\boldsymbol{ACT_{i,v,t}}$, and derived decision variables $\boldsymbol{ES_t}, \boldsymbol{E_t}, \boldsymbol{P_t}$ and $\boldsymbol{\theta_t}$, the nonlinear objective function represented in (14) maximizes the total welfare of the system. Constraint (14a) represents a set of linear constraints that represent supply-demand equilibrium, where electricity demand is a dependent variable. Constraint (14b) is same as constraint (1b), which represents the relationship between available capacity and activity. Constraint (14c to 14f) are a set of nonlinear constraints that represents the quantity demanded of the energy service, marginal price of energy service demand, the quantity of electricity produced, and the quantity of energy efficiency required, respectively, as a function of price of electricity, $\boldsymbol{PE_t}$. Constraint (14g), which is same as Constraint (1c), is a set of all other linear constraints in the ESOM. The above optimization problem finds the optimal market clearing conditions, i.e., the optimal value of variables that maximize the consumer and producer surplus.

### 3.2. Solution Methodology

Given the assumptions for the underlying demand function, the resulting model (14) is a large-scale, welfare maximization problem with a nonlinear objective function, nonlinear and linear equality constraints, and linear inequality constraints. Since this representation has non-linear terms in the objective function as well as in the constraints, it is necessary to use nonlinear optimization methods and solvers to solve it. To solve Temoa-EE+, which is implemented in Pyomo (Hart et al., 2012), we use an Interior Point Optimizer (Ipopt) (Biegler and Zavala, 2009), which is a software package for large-scale nonlinear optimization. Ipopt is written in C++, released as open-source code under the Eclipse Public License, and is designed to find solutions of mathematical optimization problems of the form:

$$\min f(X)$$
$$\text{s.t.} \, g_L \leq g(X) \leq g_U$$
$$X_L \leq X \leq X_U$$

where $f(X): R^n \to R$ is an objective function, and $g(X): R^n \to R^m$ is a set of constraint functions. The vectors $g_L$ and $g_U$ denote the lower and upper bounds on the constraints, and the vectors $X_L$ and $X_U$ are the bounds on the decision variables $X$. The functions $f(X)$ and $g(X)$ can be nonlinear and nonconvex but should be twice continuously differentiable.

Due to the nonlinear nature of the model, Ipopt only guarantees the local optimality of the solution. However, to generate insights for policy analysis, finding the global optimal solution is necessary. In order to prove the global optimality of the solution, we modify the model given in (14) by introducing the production function for $ES_t$ according to Equation (5). Hence, $ES_t$ can be written as

$$ES_t = \left( \alpha \cdot \boldsymbol{\theta}_t^{(\sigma-1)/\sigma} + (1-\alpha) \cdot \boldsymbol{E}_t^{(\sigma-1)/\sigma} \right)^{\sigma/(\sigma-1)}$$

We replace the price of energy services denoted by $P_t(q)$ in the objective function (14) by $(q/\varphi)^{1/\epsilon}$ since $P_t$ can be written as $(ES_t/\varphi_t)^{1/\epsilon}$ from Equation (14c). The resulting mathematical model is given in (15).

$$\max \sum_{t \in T} \int_{ES_t^{min}}^{\left( \alpha \cdot \boldsymbol{\theta}_t^{(\sigma-1)/\sigma} + (1-\alpha) \cdot \boldsymbol{E}_t^{(\sigma-1)/\sigma} \right)^{\sigma/(\sigma-1)}} (q/\varphi)^{1/\epsilon} \, dq - \sum_{t \in T} P\theta \cdot \boldsymbol{\theta}_t - \sum_{v \in V} \sum_{i \in I} IC_{i,v} \cdot \boldsymbol{CAP}_{i,v} \quad (15)$$

$$- \sum_{t \in T} \sum_{v \in V} \sum_{i \in I} FC_{i,v,t} \cdot \boldsymbol{CAP}_{i,v} - \sum_{t \in T} \sum_{v \in V} \sum_{i \in I} VC_{i,v,t} \cdot \boldsymbol{ACT}_{i,v,t}$$

$$\text{s.t.} \sum_{v \in V_i} \sum_{i \in I} \boldsymbol{ACT}_{i,v,t} \geq \boldsymbol{E}_t \qquad \forall\, t \in T \quad (15a)$$

$$\zeta_{i,v,t} \cdot \boldsymbol{CAP}_{i,v} \geq \boldsymbol{ACT}_{i,v,t} \qquad \forall\, t \in T_i, v \in V_i, i \in I \quad (15b)$$

$$B \cdot \boldsymbol{X} \geq b \qquad (15c)$$

We then prove that the objective function of (15) is concave, and the feasible domain is closed and convex. Since a local maxima is a global maxima for a concave function on a closed, convex feasible domain, we conclude that the solution obtained by Ipopt is, in fact, a global maximum. The proof of global optimality of a solution obtained from this nonlinear model formulation is given in Appendix A.

4. Test Case

To demonstrate the utility of the Temoa-EE+ formulation, we perform tests on a simple, hypothetical system. Imagine an island that has one diesel generator to satisfy all of its electricity demand.

For the sake of simplicity, we assume that the island has only one season and that the electricity demand is constant over the entire day. In 2020, we observe that the island has a residential lighting demand of 525 million lumen-hours, which can be satisfied with conventional lightbulbs with an efficacy of 15 lumens/watt or energy-efficient lightbulbs with an efficacy of 20 lumens/watt. Furthermore, we observe that the conventional lightbulbs consumed 16 MWh of electricity and the energy-efficient lightbulbs consumed 14.2 MWh of electricity, which at an electricity price of $0.12/kWh, cost consumers a total of $4,334 for residential lighting. For the purpose of this test system, the electricity consumption of 16 MWh from conventional lightbulbs in 2020 is considered to be the reference electricity demand, $E_t^0$, while the electricity price, considered as the reference electricity price, $PE_t^0$, is $0.12 / kWh. As the price elasticity of electricity usage for residential demand is estimated to be in the [-1, -0.1] range by Burke and Abayasekara (2018), we assume that the own-price elasticity of lighting demand, $\epsilon$, is -0.4. In 2021, we assume that the island's government has decided to provide an investment subsidy for energy-efficient residential lighting to reduce emissions from the diesel generator.

Energy efficiency can be considered as energy consumption avoided, and thus is often measured by "negawatts" (Palmer and Paul, 2015). There is fairly extensive literature examining the cost-effectiveness of energy efficiency or demand-side management programs. Common cost values in the literature (i.e., the total expense of running the program and installing equipment) as a dollar per megawatt-hour saved as a result of the program range from below $10/MWh to above $200/MWh (in real 2002 dollars) (Gillingham et al., 2009). Hence, we assume a marginal cost of energy-efficient technology within the observed range, 50 $/MWh. In this case study, the marginal cost can be interpreted as the investment cost required to switch from the pre-existing, conventional bulbs to energy-efficient ones. All the parameters required to represent the hypothetical test case are given in Table 2. Note that the fuel cost is included in the variable cost of the power plant, and the marginal damage of $CO_2$ emissions is explained in Section 5.

     16

**Table 2:** Test model parameter values

| Model parameter | Value |
| --- | --- |
| Existing time period | 2018 |
| Future time period | 2019 |
| Input commodity | Diesel |
| Output commodity | Electricity |
| Existing capacity (GW) | 0.01 |
| Investment cost ($/kW) | 1500 |
| Fixed cost ($/kW-yr) | 20 |
| Variable cost ($/kWh) | 0.25 |
| Marginal cost of meeting demand in 2018 ($/kWh) | 0.12 |
| Productivity of energy-efficient technology, $\alpha$ | 0.5714 |
| Elasticity of substitution between electricity and energy efficiency, $\sigma$ | 2.0 |
| Own price demand elasticity of energy service, $\epsilon$ | -0.4 |
| Cost of energy-efficient technology, ($/kWh) | 0.05 |
| Marginal cost of energy efficiency, $P\theta$ ($/kWh) | 0.17 |
| Residential lighting demand (million lumen-hours) | 525 |
| Marginal damage of $CO_2$ emissions, $\tau$ ($/tCO_2$) | 40 |

This simplistic, hypothetical test case is used to demonstrate the functionality of the Temoa-EE+ model, and given its simplicity, allows us to isolate and observe the tradeoff between energy supply and efficiency. We also include a slightly more complex energy system representation that includes multiple electricity supply technologies in Appendix B. This representation can be further extended to represent a more realistic energy system, with additional supply technologies and service demands, that endogenizes the tradeoff between an energy efficiency subsidy, energy consumption, and service demand levels. The results show similar behavior as the test case results presented in Section 6.

## 5. Policy Scenarios

To analyze the policy scenarios, we borrow the first-best and second-best terminology from the economics literature. Economists refer to the first-best policy as the option that gives the welfare-maximizing outcome, which is equivalent to the optimal strategy. In contrast, the second-best policy is a suboptimal strategy that is closest to the optimal strategy. We consider a Pigouvian tax (Pigou, 2017), where the tax value is set equal to the marginal external damage $\tau$ since the Pigouvian tax achieves the first-best policy outcome in the case of a single pollutant. The electricity producer must account for the additional cost associated with the emissions tax, such that the marginal cost of energy services from each generation technology, inclusive of emissions damages, is equated across sources and with energy efficiency. Ricke et al. (2018) calculate the social cost of carbon for the United States to be between 10 to 50 \$/tCO$_2$. Hence, for the test case, we choose a carbon tax within this range equal to 40 \$/tCO$_2$.

With regard to energy efficiency, we assume the island government provides a subsidy, $\beta$, to incentivize the adoption of energy efficient lighting. To compute the welfare considering a Pigouvian tax, the objective function represented by (14) is modified to include an emissions tax as given in (16) subject to constraints (14a-g).

$$\max \sum_{t \in T} \int_{ES_t^{min}}^{ES_t} \boldsymbol{P}_t(q)\, dq - \sum_{t \in T} P\theta \cdot \boldsymbol{\theta}_t - \sum_{v \in V_i} \sum_{i \in I} IC_{i,v} \cdot \boldsymbol{CAP}_{i,v} - \sum_{t \in T_i} \sum_{v \in V_i} \sum_{i \in I} FC_{i,v,t} \cdot \boldsymbol{CAP}_{i,v} \quad (16)$$

$$- \sum_{t \in T_i} \sum_{v \in V_i} \sum_{i \in I} VC_{i,v,t} \cdot \boldsymbol{ACT}_{i,v,t} - \sum_{t \in T_i} \sum_{v \in V_i} \sum_{i \in I} \tau\, \gamma_{i,vt} \cdot \boldsymbol{ACT}_{i,v,t}$$

The parameter $\gamma_{i,vt}$ in (16) is the emission activity of technology $i$ with vintage $v$ in time period $t$. Thus, $\sum_{t \in T_i} \sum_{v \in V_i} \sum_{i \in I} \tau \cdot \gamma_{i,vt} \cdot \boldsymbol{ACT}_{i,v,t}$ represents the total emissions subject to the Pigouvian tax. To compute the welfare considering an efficiency subsidy, the Temoa-EE+ formulation given in (14) is modified to include the subsidy, as given in (17):

$$\max \sum_{t \in T} \int_{ES_t^{min}}^{ES_t} \boldsymbol{P}_t(q)\, dq - (1 - \beta) \cdot \sum_{t \in T} P\theta \cdot \boldsymbol{\theta}_t - \sum_{v \in V_i} \sum_{i \in I} IC_{i,v} \cdot \boldsymbol{CAP}_{i,v} - \sum_{t \in T_i} \sum_{v \in V_i} \sum_{i \in I} FC_{i,v,t} \cdot \boldsymbol{CAP}_{i,v} \quad (17)$$

$$- \sum_{t \in T_i} \sum_{v \in V_i} \sum_{i \in I} VC_{i,v,t} \cdot \boldsymbol{ACT}_{i,v,t}$$

s.t. $\sum_{v \in V_i} \sum_{i \in I} \boldsymbol{ACT}_{i,v,t} \geq \boldsymbol{E}_t$ $\forall\, t \in T$ (17a)

$\zeta_{i,v,t} \cdot \boldsymbol{CAP}_{i,v} \geq \boldsymbol{ACT}_{i,v,t}$ $\forall\, t \in T_i, v \in V_i, i \in I$ (17b)

$\boldsymbol{ES}_t = \varphi_t \cdot \boldsymbol{P}_t^\epsilon$ $\forall\, t \in T$ (17c)

$\boldsymbol{P}_t = (\alpha^\sigma \cdot ((1 - \beta) \cdot P\theta)^{1-\sigma} + (1 - \alpha)^\sigma \cdot \boldsymbol{PE}_t^{1-\sigma})^{1/(1-\sigma)}$ $\forall\, t \in T$ (17d)

$\boldsymbol{E}_t = \varphi_t \cdot (1 - \alpha)^\sigma \cdot \boldsymbol{PE}_t^{-\sigma}\, (\alpha^\sigma \cdot ((1 - \beta) \cdot P\theta)^{1-\sigma} + (1 - \alpha)^\sigma \cdot \boldsymbol{PE}_t^{1-\sigma})^{(\epsilon+\sigma)/(1-\sigma)}$ $\forall\, t \in T$ (17e)

$\boldsymbol{\theta}_t = \varphi_t \cdot \alpha^\sigma \cdot ((1 - \beta) \cdot P\theta)^{-\sigma}$
$\cdot (\alpha^\sigma \cdot ((1 - \beta) \cdot P\theta)^{1-\sigma} + (1 - \alpha)^\sigma \cdot \boldsymbol{PE}_t^{1-\sigma})^{(\epsilon+\sigma)/(1-\sigma)}$ $\forall\, t \in T$ (17f)

$B\boldsymbol{X} \geq b$ (17g)

Where, $\beta$ represents the energy efficiency subsidy. Thus, fixing $\beta$ to 0.3 in (17) is equivalent to assuming that the price of energy efficiency is 30% lower than the base value. The subsidy encourages buyers to invest in energy efficient lighting.

Conceptually, the carbon tax policy and efficiency policy are shown in Figure 3. We add a new actor to Figure 1 – a government – that can issue the emissions tax and energy efficiency subsidy. Since energy efficiency and electricity are substitute goods, the subsidy decreases electricity demand, which in turn reduces emissions.

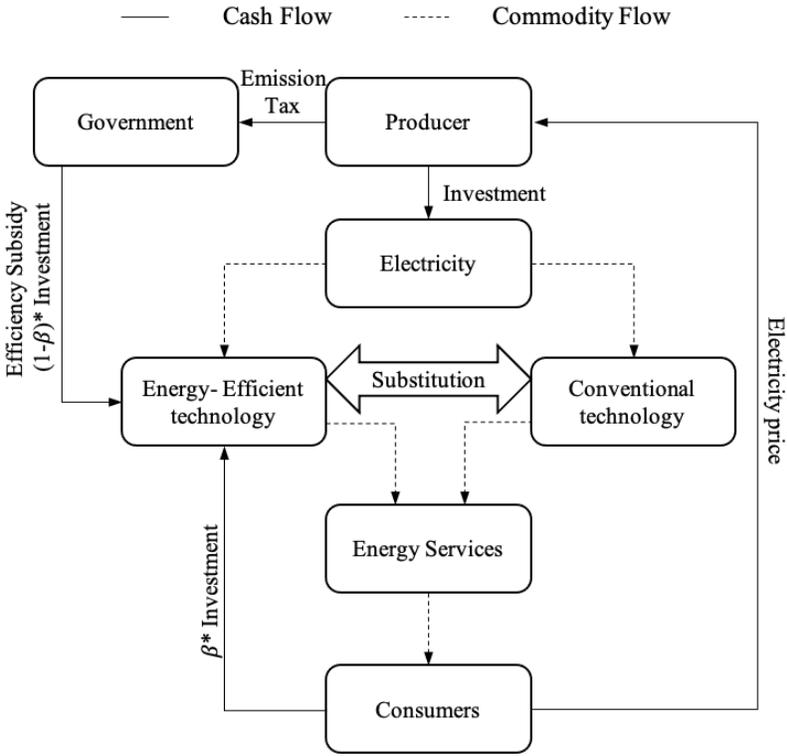

**Figure 3:** Conceptual cash and commodity flows associated with the representation of energy efficiency in Temoa-EE+. Note that this representation includes a government that can levy a carbon tax and subsidize investments in energy efficiency.

Fell et al. (2017) proves that the first best allocation, i.e., the welfare value associated with the carbon tax policy obtained by solving (16) and subject to constraints (14a-g), cannot be achieved with an efficiency subsidy unless the energy service demand is fixed in the absence of potential capacity expansion. Since the quantity of energy service demand is price responsive, i.e., elastic, we cannot achieve the first-best allocation with an efficiency subsidy. However, with an optimal choice of subsidy, we can achieve the second-best allocation, i.e., the welfare value that is closest to the one obtained with the carbon tax. In the following analysis, the 'no policy' case represents the solution to Temoa-EE+ as given in (14). Solving the model with the efficiency subsidy is equivalent to solving the mathematical model given in (17) where $0 < \beta < 1$. The percentage welfare recovered, $\%W$, from the efficiency subsidy compared to the carbon tax policy is given as:

$$\%W = \frac{(W_{ES} - W_{NP})}{(W_{ET} - W_{NP})} \qquad (18)$$

Where, $W_{ES}$ and $W_{ET}$ represent welfare from efficiency subsidy and emission tax policy, and $W_{NP}$ represents welfare from no policy scenario. When comparing the change in welfare associated with different policies, we assume that the welfare from the carbon tax policy is equivalent to the optimal objective function of (16). Now, let $A$ be the optimal objective function value of the Temoa-EE+ model given in (17). The objective function associated with the efficiency subsidy obtained by solving (17) does not include the carbon tax or the cost of the subsidy offered by the government. Thus, the net welfare from the efficiency subsidy policy must be calculated ex-post, taking into account the cost of damage equal to the Pigouvian tax and the efficiency subsidy:

$$W_{ES}^* = A - \sum_{t \in T_i} \sum_{v \in V_i} \sum_{i \in I} \tau \cdot \gamma_{i,vt} \cdot \mathbf{ACT}_{i,v,t}^* - \beta \cdot \sum_{t \in T} P\theta \cdot \boldsymbol{\theta}_t^* \qquad (19)$$

Where, $W_{ES}^*$ represents actual welfare from efficiency subsidy policy, and $\mathbf{ACT}_{i,v,t}^*$ and $\boldsymbol{\theta}_t^*$ are the optimal values of the variables obtained by solving (17).

6. **Results and Discussion**

In our illustrative case study, meeting the 2020 residential lighting demand of 525 million lumen-hours using only conventional lightbulbs would consume 35 MWh of electricity and cost $4,200. In contrast, using only energy-efficient light bulbs would consume 26.25 MWh of electricity and cost $4,462, which includes both the cost of electricity and the cost to upgrade, represented by the marginal cost of energy efficiency. Under these conditions, the traditional ESOM would choose the least-cost option and use only conventional lightbulbs to satisfy residential lighting demand. Moreover, a 30% subsidy for the energy-efficient bulbs would force the traditional ESOM to flip its decision and install only energy-efficient bulbs at a total cost of $4,068 to consumers. The traditional ESOM would not consider the effect of the subsidy on electricity consumption. By focusing exclusively on relative cost, traditional, least-cost ESOMs often produce knife-edge solutions that involve a wholesale switch from one technology to another. Typical kluges to address this model behavior include imposing share constraints that force the model to use both bulb technologies, or adding a technology-specific discount rate (i.e., hurdle rate) that makes the efficient bulbs more expensive to the model, thereby suppressing their uptake. Neither of these approaches has a strong theoretical or empirical grounding, but rather rely on subjective modeler judgement.

By contrast, the Temoa-EE+ model outlined above provides a way to capture the tradeoff between these two technologies in a way that is consistent with microeconomic theory. Given a 30% efficiency subsidy, the model would produce a 7.5% decrease in electricity consumption from conventional light bulbs (14.8 MWh), a 10.6% increase in electricity consumption from energy-efficient light bulbs (15.7 MWh), and a 3.8% increase in residential lighting demand (545 million lumen-hours), relative to the observed values provided in Section 4. This approach results in a 3.6% difference in cost to consumers ($4216), and the knife-edge behavior of the traditional, least-cost ESOMs is successfully avoided.

We know that an increase in electricity price will simultaneously decrease electricity demand and increase both the demand for energy efficiency (e.g., the more efficient bulbs) and the price of energy services. Since energy efficiency and electricity are substitute goods, subsidizing energy efficiency will decrease the investment in electricity production. Hence, for a given electricity price, an increase in the efficiency subsidy will increase the quantity of energy efficiency demanded and the energy service demand, while decreasing the quantity of electricity demanded. In (14), the decision variables, $\boldsymbol{ES_t}, \boldsymbol{E_t}, \boldsymbol{P_t}$ and $\boldsymbol{\theta_t}$ are derived from the electricity price, $\boldsymbol{PE_t}$. To demonstrate the relation between the derived variables and $\boldsymbol{PE_t}$, we compute Equations 14(c), (d), (e), and (f) by varying $\boldsymbol{PE_t}$ from 0.05 to 0.4 $/kWh. Figure 4 includes the variation in electricity demand, energy efficiency, and energy services as a function of the electricity price and subsidy level.

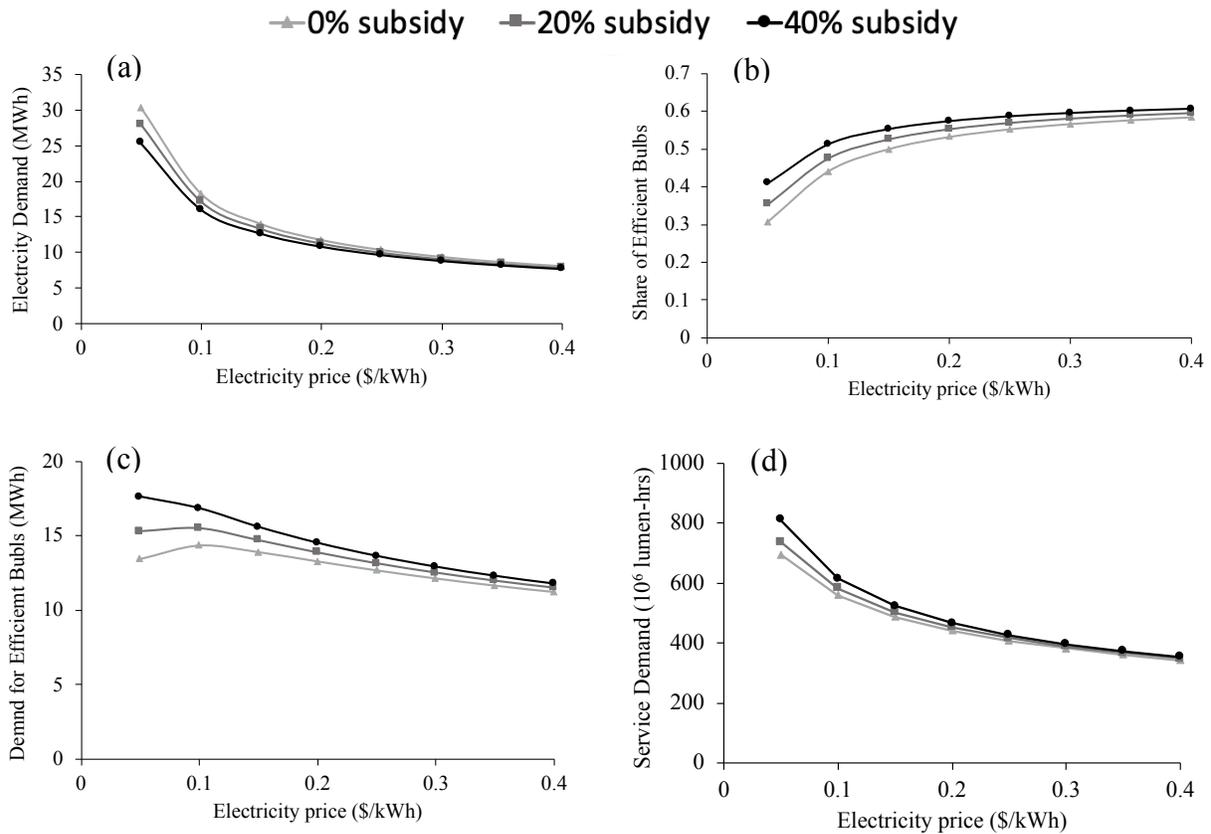

**Figure 4:** For a given electricity price, (a) electricity consumption associated with conventional technology decreases with an increasing efficiency subsidy, (b) the fraction of residential lighting demand satisfied with energy-efficient technology increases with an increasing efficiency subsidy, (c) electricity consumption associated with energy-efficient bulbs increases with an increasing efficiency subsidy, and (d) residential lighting demand increases with increasing efficiency subsidy.

In Figure 4, we vary the efficiency subsidy over a large range (0-40%) and observe that the efficiency subsidy affects the rate at which the electricity demand decreases with an increase in the electricity price. By varying input parameters in Temoa-EE+, particularly the parameters in Equation 14c-f, we can incorporate various consumer behaviors as a function of the electricity price. One aspect of the current model worth noting is the inclusion of a rebound effect where, over the long run, the efficiency subsidy induces a decline in the price of energy services, which leads to an increase in energy service consumption. However, we do not directly isolate the rebound effect in this analysis. Note that the price of energy services does not have a real world analogue since we do not directly pay for energy services (e.g., lumens of light). The price of energy services can be thought of as a function of the price of electricity and the price of energy efficiency. If the electricity or energy efficiency price increases, it leads to an increase in the price of energy services.

Figure 5 illustrates the welfare gain from varying levels of the efficiency subsidy compared to the welfare gain from the carbon tax. One of the effects of an efficiency subsidy is a reduction in electricity demand, which reduces emissions. As a result, investing in energy efficiency reduces the damage associated with emissions. The results indicate that an efficiency subsidy of 6% achieves maximum welfare, which is 38% of the welfare gain from the carbon tax policy. Beyond a 12% efficiency subsidy, the cost savings from the emissions reduction is less than the combination of energy efficiency expenditures and cost of damages, leading to a negative welfare gain.

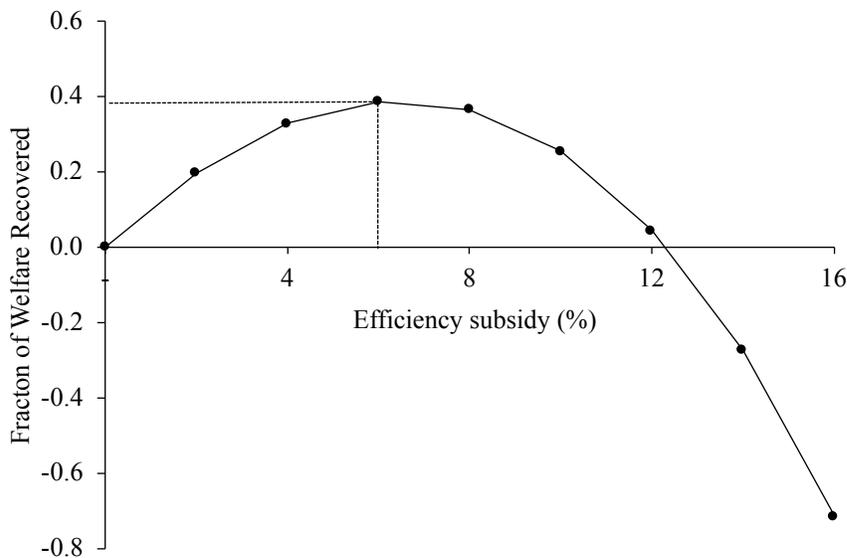

**Figure 5:** Welfare gain as a function of the efficiency subsidy. The gain is expressed as the fraction of welfare gain with a Pigouvian carbon tax set at 40 $/ton of $CO_2$. As indicated by the dotted lines, an efficient lightbulb subsidy of 6% recovers the maximum amount of welfare (nearly 40%) relative to the tax.

Figure 6 presents price and quantity results for a range of efficiency subsidies. In Figure 6, emissions are 8% higher under the no policy scenario (i.e., no efficiency subsidy) and gradually decrease with an increasing efficiency subsidy due to a decrease in the quantity of electricity demanded. According to Equation (14d), an increase in the efficiency subsidy $\beta$ reduces the price of lighting services, $P_t$. Likewise, Equation (14c) indicates that a decrease in the price of lighting service leads to an increase in the demand for lighting service, $ES_t$. Thus, the efficiency subsidy increases demand for the energy-efficient bulbs and a decrease in the unit price of lighting service demand, which in turn increases the quantity lighting service demand. In addition, an increasing energy efficiency subsidy decreases the electricity demand. As a result, emissions from the diesel generator decrease.

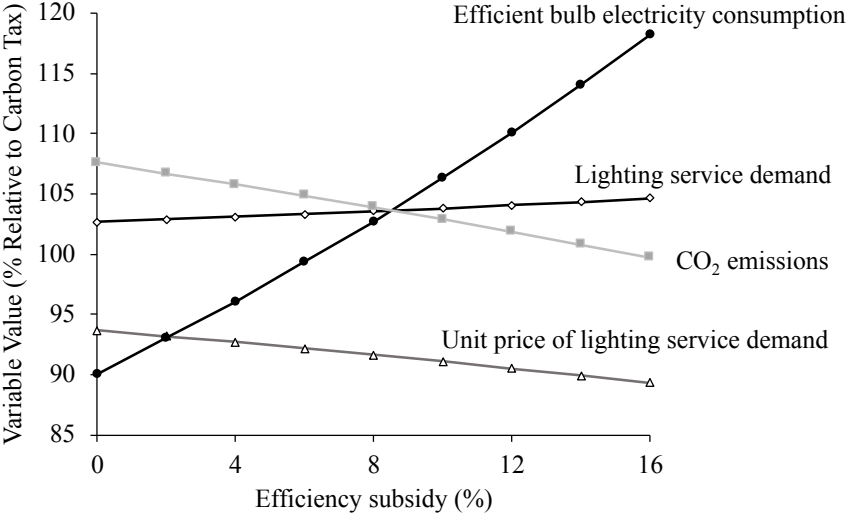

**Figure 6:** Prices and quantity demanded as a function of the efficiency subsidy. Variable values are relative to the values under a 40 $/ton emission tax policy.

In this proposed Temoa-EE+ formulation, it is important to consider the inherent uncertainty in the assumed parameter values. The substitution parameter, $\sigma$, represents the consumer's willingness to invest in energy-efficient technologies for a marginal increase in the electricity price. The higher the willingness to substitute electricity with energy efficiency, the higher the value of the substitution parameter, $\sigma$. The price elasticity parameter, $\epsilon$, denotes the importance of energy services for the consumer. Higher elasticity values imply that the consumer is more willing to reduce energy service consumption if it is marginally more expensive. Moreover, the productivity parameter, $\alpha$, represents the consumer's perspective on the energy services obtained from energy-efficient technologies. If the consumer views investing in energy efficiency as a superior option, then the productivity of energy efficiency is higher, leading to a higher value of $\alpha$. Such consumer behaviors are inherently uncertain, and they can vary over a broad range for different groups of consumers depending on their social and economic status. Also, carbon taxes vary worldwide from 0 $/ton to 130 $/ton of carbon (World Bank and Ecofys, 2018). In this analysis, we assume a scalar value for energy efficiency cost. However, in reality, the cost of energy efficiency, $P\theta_t$, can vary over a wide range depending on the type of energy-efficient technology. We perform a sensitivity analysis to evaluate the overall impact of these system parameters on total welfare.

We consider the model parameter values from the test case given in Table 2 as the base case values. We vary the system parameters mentioned above ±50% from the base case values given in Table 2, except for the elasticity of substitution, $\sigma$. Decreasing the elasticity of substitution by 50% from the base value leads to model infeasibility since the CES production function used for this analysis is undefined for $\sigma = 1$. Therefore, we only present results for a 50% increase in the elasticity of substitution parameter. The sensitivity analysis on the productivity of energy efficiency, $\alpha$, suggests that for very low ($\alpha < 0.2$) or very high ($\alpha > 0.7$) values of energy efficiency productivity, the relative welfare gain with an efficiency subsidy is not significant. The relatively high productivity of energy efficiency reduces the need to subsidize it, while the relatively low productivity reduces the effect of the subsidy. Figure 7 below shows the effect of the four uncertain parameters on the relative welfare gain from the efficiency subsidy.

Figure 7(a) suggests that a lower cost for energy efficiency leads to higher welfare recovered. A higher degree of substitution between electricity and energy efficiency produces higher welfare, as shown in Figure 7(b). Figure 7(c) suggests that a lower own-price elasticity of end-use energy service demand increases the welfare recovered at a given efficiency subsidy. Figure 7(d) suggests that a higher efficiency subsidy is needed to recover the maximum welfare at a higher carbon tax.

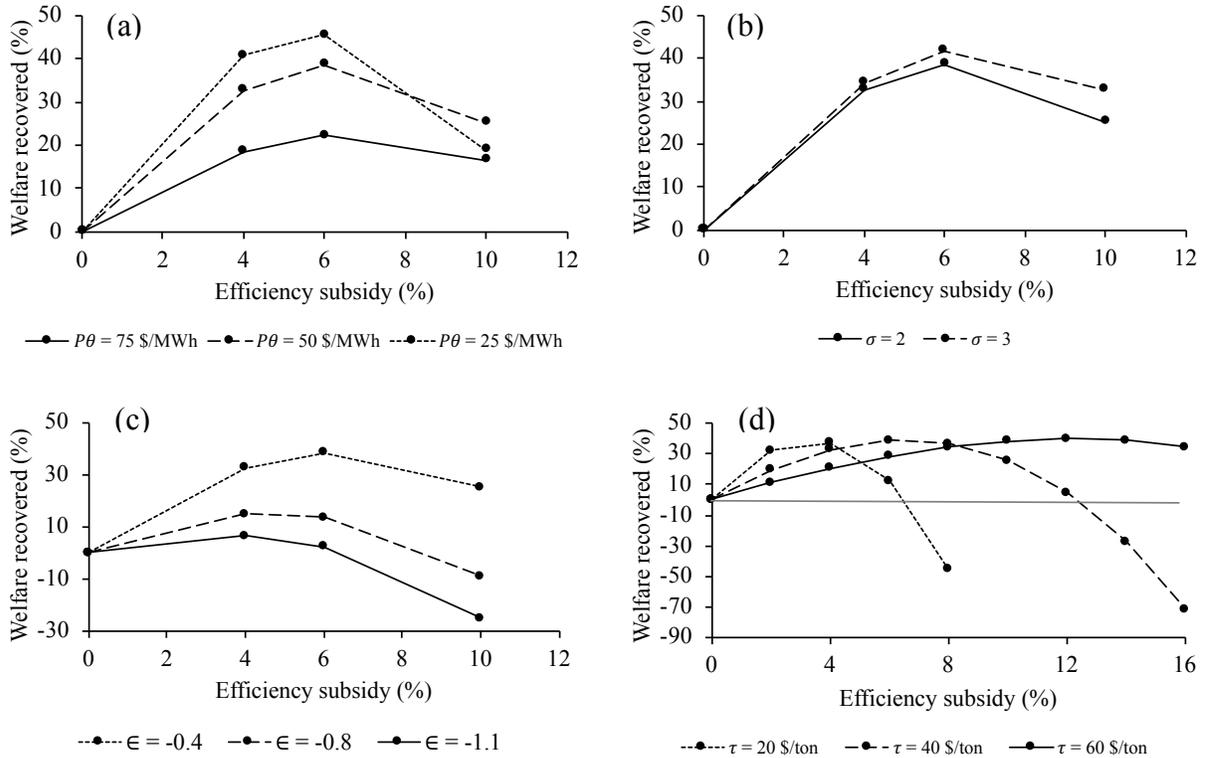

**Figure 7:** Effect of uncertain model parameters on the welfare recovered through an energy efficiency subsidy compared with the Pigouvian tax. Uncertain parameters are (a) energy efficiency cost (P$\theta$), (b) substitutability between electricity and energy efficiency ($\sigma$), (c) own-price elasticity of energy service demand ($\epsilon$), and (d) carbon tax ($\tau$).

The purpose of a carbon tax and efficiency subsidy is to reduce emissions by discouraging electricity usage. The former achieves emissions reductions by increasing the price of electricity while the latter does so by decreasing the price of energy efficiency. The relative welfare gain from an efficiency subsidy to that of a carbon tax depends on the efficiency-induced savings on emissions-induced damage versus the increased expenditure on energy efficiency. When the savings associated with avoided emissions-induced damage is greater than the expenditure on energy efficiency, the relative welfare recovered from the efficiency subsidy is positive.

To analyze Figure 7(a) further, note that electricity and energy efficiency are substitute goods. As a result, the effect of a marginal reduction in energy efficiency cost on the quantity of electricity demanded is higher when the energy efficiency cost is lower. In other words, the reduction in electricity generation is higher at a 10% subsidy when $P\theta$ is 25 $/MWh than when $P\theta$ is 75 $/MWh. Hence, relative welfare recovered from an efficiency subsidy increases with a lower energy efficiency cost. Similar logic can be applied to Figure 7(b). An increase in the substitutability of energy efficiency, $\sigma$, increases the reduction in electricity generation for a marginal decrease in $P\theta$. Consequently, a 10% efficiency subsidy produces a larger reduction in electricity production for a higher value of $\sigma$, leading to higher relative welfare recovered. As for Figure 7(c), higher price elasticity of demand implies that an increase in the quantity demanded of a good is higher for a given marginal reduction in the price of a good. Therefore, at a 10% energy efficiency subsidy, the higher elasticity values produce a larger increase in energy service demand, and consequently, energy efficiency demand, compared to lower elasticity values. The rate of increase in energy efficiency demand or decrease in electricity demand depends on other model parameters, such as the productivity of energy efficiency, $\alpha$. However, for the set of parameters given in Table 2, Figure 7(c) suggests that an increase in energy efficiency expenditure is greater than the efficiency subsidy-induced savings on emissions-induced damage for more elastic energy service demands. As a result, the relative welfare gain from the efficiency subsidy decreases as the elasticity of energy service demand increases. Figure 7(d) suggests that for a higher carbon tax, we need a higher efficiency subsidy to recover the same amount of relative welfare. Higher carbon taxes lead to higher emission reductions. To achieve an equivalent emissions reduction, we need lower electricity demand and higher energy efficiency demand. Such an outcome can be achieved when the cost of energy efficiency is low or equivalently, the subsidy for energy efficiency is high. Note that in reality, we cannot fully satisfy the end-use energy service demand by energy efficiency since that would imply zero energy consumption.

## 7. Conclusions and Future Work

The primary goal of this work is to introduce consumer behavior in energy system optimization models (ESOMs) and formulate it in a way that is consistent with microeconomic theory. To do so, we restructured an existing, open-source ESOM to represent the tradeoff between energy efficiency and electricity supply. We apply the methodology to perform policy analysis for a hypothetical test case. We point out the differences between a traditional ESOM where we define energy efficiency through separate technologies and Temoa-EE+, which can explicitly model the substitutability between electricity and energy efficiency in the form of a production function. The secondary goal is to analyze the effect of uncertain consumer behavior on system welfare. Substitutability between electricity and energy efficiency, the productivity of energy efficiency in satisfying energy service demand, and the price responsiveness of energy service demand can be used to tune consumer response. By varying these model parameters, we can potentially incorporate a wide range of consumer behavior related to energy consumption into traditional ESOMs. For example, the productivity parameter, $\alpha$, represents the consumers' view on energy-efficient technologies relative to traditional supply-side generation. If consumers view energy-efficient technology as superior to consuming more electricity, then energy efficiency will be very productive in generating energy services, and the value of $\alpha$ should be higher.

Although ESOMs can benefit from the introduction of a methodology that considers consumer behavior, it has some limitations. One limitation is the narrow literature on quantifying consumer behavior related to the uptake of energy-efficient technologies: determining the appropriate value of the substitution parameter, $\sigma$, the productivity of energy efficiency, $\alpha$, and the price elasticity of energy services, $\epsilon$, is a challenging task. However, performing sensitivity analysis on these parameters can provide valuable insights regarding the effect of an efficiency subsidy on overall system behavior. In addition, we emphasize that the enhanced formulation presented here allows for demand and price adjustments across model scenarios that are internally consistent and align with microeconomic theory. As with all model results, insights should be drawn from a wide range of scenarios rather than a single, specific numerical result.

Another limitation of Temoa-EE+ arises from its highly nonlinear nature, which limits the size of the problem that can be solved within a reasonable computational time. Moreover, we have to rely on nonlinear solvers such as Ipopt for determining the global optimality of the resulting solution. Despite these limitations, the model provides a theoretically consistent methodology to consider some of the consumer behaviors that traditional ESOMs do not.

This work can be extended in several ways. One could apply the same methodology to model the substitution effect between different fuels or different technologies within ESOMs. For example, one could model a substitution effect between electric and gasoline vehicles in the transportation sector or the substitution effect between solar photovoltaics and natural gas generators when investment in solar is subsidized. In addition, a time index for the subsidy level $\beta$ would be helpful, since the subsidies can vary over time. Moreover, incorporating a supply curve for energy-efficient technology options instead of assuming a scalar cost value will also produce more realistic results.

**Acknowledgements**

The authors acknowledge the financial support provided by the Collaborative REsearch of Decentralization, ElectrificatioN, Communications and Economics (CREDENCE) project under the NSF grant 081212.

# Appendix A

**Proving global optimality of the IPOPT solution:**

Simplified form of the Temoa-EE+ model in (14) can be given as follows:

$$\max f(E_t, \theta_t, CAP_{i,v}, ACT_{i,v,t}) \tag{1}$$

$$= \sum_{t \in T} \int_{ES_t^{min}}^{ES_t} (q/\varphi)^{1/\epsilon} dq - \beta \sum_{t \in T} P\theta \cdot \theta_t - \sum_{v \in V_i} \sum_{i \in I} IC_{i,v} \cdot CAP_{i,v}$$

$$- \sum_{t \in T_i} \sum_{v \in V_i} \sum_{i \in I} FC_{i,v,t} \cdot CAP_{i,v} - \sum_{t \in T_i} \sum_{v \in V_i} \sum_{i \in I} VC_{i,v,t} \cdot ACT_{i,v,t}$$

$$\text{s.t.} \sum_{v \in V_i} \sum_{i \in I} ACT_{i,v,t} \geq E_t \qquad \forall\, t \in T \tag{1a}$$

$$\zeta_{i,v,t} \cdot CAP_{i,v} \geq ACT_{i,v,t} \qquad \forall\, t \in T_i, v \in V_i, i \in I \tag{1b}$$

$$B \cdot X \geq b \tag{1c}$$

Where, $ES_t$ is defined as a production function of $\theta_t$ and $E_t$, given in Equation 5. Hence, $ES_t$ is equivalent to $\left(\alpha \cdot \theta_t^{(\sigma-1)/\sigma} + (1-\alpha) \cdot E_t^{(\sigma-1)/\sigma}\right)^{\sigma/\sigma-1}$. Assume that the above problem has $n$ decision variables and $m$ constraints. The proofs of the propositions 1 to 5 are given in Convex analysis and monotone operator theory in Hilbert spaces, chapter 8 (Bauschke and Combettes, 2017).

**Definition 1:** For the $n \times n$ matrix $A$, the $k^{th}$ order principal submatrix is obtained by deleting the last $n - k$ rows and columns of $A$. The determinant of this matrix is called leading principal minor of $A$.

We denote the $k^{th}$ order leading principal submatrix of $A$ by $A_k$ and the $k^{th}$ order leading principal minor by $|A_k|$.

**Proposition 1:** The matrix $A$ is negative semidefinite if and only if every principal minor of odd order is $\leq 0$ and every principal minor of even order is $\geq 0$.

**Proposition 2:** Let $f$ be a twice differentiable function on $Z$ such that $Z \in R^n$ and $x^*$ is an interior point of $Z$. Then, $f$ is concave if and only if Hessian matrix of $f$, $H_f(x^*)$, is negative semidefinite at all $x^* \in Z$.

**Proposition 3:** Let $f_1$ and $f_2$ are concave functions then $f_1 + f_2$ is also a concave function

**Proposition 4:** Feasible region $S$ where, $S = \{X \in R^n, X | AX \geq b\}$ is convex where, $X, X \in R^n$, is a vector of decision variable, $A$ is the constraint coefficient matrix and $b$ is the right-hand side of the constraint.

**Proposition 5:** Let $f$ be a concave function on a convex feasible domain $S \in R^n$ and $x^*$ be an interior point of $S$. If $x^*$ is a local maximum of a $f$ then $x^*$ is also a global maximum.

For our analysis following table represents the ranges of the parameters used for sensitivity analysis. Note that open interval denotes that the boundary values are not included.

**Appendix Table 1:** Valid range of the model parameters

| Parameter | Range |
|---|---|
| $\alpha$ | $(0,1)$ |
| $\beta$ | $(0,1)$ |
| $\epsilon$ | $(-\infty, 0)$ |
| $\sigma$ | $(1, \infty)$ |
| $P\theta$ | $(0, \infty)$ |

For the simplification purpose, we drop the summation over time period since from Proposition 3, we know that summation of concave functions is concave. For a given $\tilde{t}$, where $\tilde{t} \in T$, after simplifying the integral and ignoring the summation over time period, $f(E_t, \theta_t, CAP_{i,v}, ACT_{i,v,t})$ can be given as

$$f(E_{\tilde{t}}, \theta_{\tilde{t}}, CAP_{i,v}, ACT_{i,v,\tilde{t}}) \tag{2}$$

$$= \frac{1}{1+\epsilon} \cdot \left( \epsilon \cdot \varphi_{\tilde{t}} \cdot \left( \frac{\left( \alpha \cdot \theta_{\tilde{t}}^{(\sigma-1)/\sigma} + (1-\alpha) \cdot E_{\tilde{t}}^{(\sigma-1)/\sigma} \right)^{\sigma/(\sigma-1)}}{\varphi_{\tilde{t}}} \right)^{1+1/\epsilon} - \epsilon \cdot \varphi_{\tilde{t}} \cdot \left( \frac{ES_{\tilde{t}}^{min}}{\varphi_{\tilde{t}}} \right)^{1+1/\epsilon} \right)$$

$$- \beta \cdot P\theta \cdot \theta_{\tilde{t}} - \sum_{v \in V_i} \sum_{i \in I} IC_{i,v} \cdot CAP_{i,v} - \sum_{v \in V_i} \sum_{i \in I} FC_{i,v,\tilde{t}} \cdot CAP_{i,v} - \sum_{v \in V_i} \sum_{i \in I} VC_{i,v,\tilde{t}} \cdot ACT_{i,v,\tilde{t}}$$

**Proposition 6:** The function $f(E_{\tilde{t}}, \theta_{\tilde{t}}, CAP_{i,v}, ACT_{i,v,\tilde{t}})$ given in (2) is concave

We are ignoring the constant term in the objective function. The Hessian matrix, $H_f(E_{\tilde{t}}, \theta_{\tilde{t}}, CAP_{i,v}, ACT_{i,v,\tilde{t}})$ is computed as follows

$$H_f(E_{\tilde{t}}, \theta_{\tilde{t}}, CAP_{i,v}, ACT_{i,v,\tilde{t}}) = \begin{bmatrix} \frac{\partial^2 f}{\partial E_{\tilde{t}}^2} & \frac{\partial^2 f}{\partial E_{\tilde{t}} \partial \theta_{\tilde{t}}} & \frac{\partial^2 f}{\partial E_{\tilde{t}} \partial CAP_{i,v}} & \frac{\partial^2 f}{\partial E_{\tilde{t}} \partial ACT_{i,v,\tilde{t}}} \\ \frac{\partial^2 f}{\partial \theta_{\tilde{t}} \partial E_{\tilde{t}}} & \frac{\partial^2 f}{\partial \theta_{\tilde{t}}^2} & \frac{\partial^2 f}{\partial \theta_{\tilde{t}} \partial CAP_{i,v}} & \frac{\partial^2 f}{\partial \theta_{\tilde{t}} \partial ACT_{i,v,\tilde{t}}} \\ \frac{\partial^2 f}{\partial CAP_{i,v} \partial E_{\tilde{t}}} & \frac{\partial^2 f}{\partial CAP_{i,v} \partial \theta_{\tilde{t}}} & \frac{\partial^2 f}{\partial CAP_{i,v}^2} & \frac{\partial^2 f}{\partial CAP_{i,v} \partial ACT_{i,v,\tilde{t}}} \\ \frac{\partial^2 f}{\partial ACT_{i,v,\tilde{t}} \partial E_{\tilde{t}}} & \frac{\partial^2 f}{\partial ACT_{i,v,\tilde{t}} \partial \theta_{\tilde{t}}} & \frac{\partial^2 f}{\partial ACT_{i,v,\tilde{t}} \partial CAP_{i,v}} & \frac{\partial^2 f}{\partial ACT_{i,v,\tilde{t}}^2} \end{bmatrix}$$

Where, the partial derivative with respect to $CAP_{i,v}$ is equivalent to taking a partial derivative with respect to $CAP_{\tilde{i},\tilde{v}}, \forall \tilde{i} \in I, \tilde{v} \in V_i$. Hence, the dimension of the $H_f(E_{\tilde{t}}, \theta_{\tilde{t}}, CAP_{i,v}, ACT_{i,v,\tilde{t}})$ is $[(2 + IV_i + IV_i) \times (2 + IV_i + IV_i)]$. The leading principal minor $|A_1|$ which is equivalent to $\left| \frac{\partial^2 f}{\partial E_{\tilde{t}}^2} \right|$ is given by (3):

$$\frac{(-1+\alpha) \cdot \boldsymbol{\theta}_{\tilde{t}}^{\frac{1}{\sigma}} \cdot \varphi_{\tilde{t}} \cdot \left(\alpha \cdot \boldsymbol{E}_{\tilde{t}}^{\frac{1}{\sigma}} \cdot \epsilon \cdot \boldsymbol{\theta}_{\tilde{t}} + (-1+\alpha) \cdot \boldsymbol{E}_{\tilde{t}} \cdot \sigma \cdot \boldsymbol{\theta}_{\tilde{t}}^{\frac{1}{\sigma}}\right) \cdot \left(\frac{\left((1-\alpha) \cdot \boldsymbol{E}_{\tilde{t}}^{\frac{-1+\sigma}{\sigma}} + \alpha \cdot \boldsymbol{\theta}_{\tilde{t}}^{\frac{-1+\sigma}{\sigma}}\right)^{\frac{\sigma}{-1+\sigma}}}{\varphi_{\tilde{t}}}\right)^{1+\frac{1}{\epsilon}}}{\epsilon \cdot \sigma \cdot \boldsymbol{E}_{\tilde{t}} \cdot \left(\alpha \cdot \boldsymbol{E}_{\tilde{t}}^{\frac{1}{\sigma}} \cdot \boldsymbol{\theta}_{\tilde{t}} + (1-\alpha) \cdot \boldsymbol{E}_{\tilde{t}} \cdot \boldsymbol{\theta}_{\tilde{t}}^{\frac{1}{\sigma}}\right)^2} \quad (3)$$

Assume that

$$a = (-1+\alpha) \cdot \boldsymbol{\theta}_{\tilde{t}}^{\frac{1}{\sigma}} \cdot \varphi_{\tilde{t}}$$

$$b = \left(\alpha \cdot \boldsymbol{E}_{\tilde{t}}^{\frac{1}{\sigma}} \cdot \epsilon \cdot \boldsymbol{\theta}_{\tilde{t}} + (-1+\alpha) \cdot \boldsymbol{E}_{\tilde{t}} \cdot \sigma \cdot \boldsymbol{\theta}_{\tilde{t}}^{\frac{1}{\sigma}}\right)$$

$$c = \left(\frac{\left((1-\alpha) \cdot \boldsymbol{E}_{\tilde{t}}^{\frac{-1+\sigma}{\sigma}} + \alpha \cdot \boldsymbol{\theta}_{\tilde{t}}^{\frac{-1+\sigma}{\sigma}}\right)^{\frac{\sigma}{-1+\sigma}}}{\varphi_{\tilde{t}}}\right)^{1+\frac{1}{\epsilon}}$$

$$d = \epsilon \cdot \sigma \cdot \boldsymbol{E}_{\tilde{t}} \cdot \left(\alpha \cdot \boldsymbol{E}_{\tilde{t}}^{\frac{1}{\sigma}} \cdot \boldsymbol{\theta}_{\tilde{t}} + (1-\alpha) \cdot \boldsymbol{E}_{\tilde{t}} \cdot \boldsymbol{\theta}_{\tilde{t}}^{\frac{1}{\sigma}}\right)^2$$

Hence, the Expression (3) can be written as $(abc)/d$. We compute $\varphi_{\tilde{t}}$ in equation (13). Since the demand of electricity $\boldsymbol{E}_{\tilde{t}}^0$ and price of electricity $\boldsymbol{PE}_{\tilde{t}}^0$ are non-negative decision variables, $\varphi_{\tilde{t}}$ is always positive. The decision variables $\boldsymbol{\theta}_{\tilde{t}}$ and $\boldsymbol{E}_{\tilde{t}}$ are non-negative. Hence for the assumed parameter values given in Table 1, $a, b$ and $d$ are always negative while $c$ is always positive. Hence, $(abc)/d$ is always negative, i.e., is the first principal minor is always negative.

The leading principal minor $|A_2|$ where,

$$A_2 = \begin{bmatrix} \frac{\partial^2 f}{\partial E_{\tilde{t}}^2} & \frac{\partial^2 f}{\partial E_{\tilde{t}} \boldsymbol{\theta}_{\tilde{t}}} \\ \frac{\partial^2 f}{\partial \boldsymbol{\theta}_{\tilde{t}} E_{\tilde{t}}} & \frac{\partial^2 f}{\partial \boldsymbol{\theta}_{\tilde{t}}^2} \end{bmatrix}$$

is given as (4):

$$\frac{(-1+\alpha) \cdot \alpha \cdot E_{\tilde{t}}^{-1+\frac{1}{\sigma}} \cdot \boldsymbol{\theta}_{\tilde{t}}^{-1+\frac{1}{\sigma}} \cdot \left((1-\alpha) \cdot E_{\tilde{t}}^{\frac{-1+\sigma}{\sigma}} + \alpha \cdot \boldsymbol{\theta}_{\tilde{t}}^{\frac{-1+\sigma}{\sigma}}\right)^{\frac{2\sigma}{-1+\sigma}} \cdot \left(\frac{\left((1-\alpha) \cdot E_{\tilde{t}}^{\frac{-1+\sigma}{\sigma}} + \alpha \cdot \boldsymbol{\theta}_{\tilde{t}}^{\frac{-1+\sigma}{\sigma}}\right)^{\frac{\sigma}{-1+\sigma}}}{\varphi_{\tilde{t}}}\right)^{2/\epsilon}}{\epsilon \cdot \sigma \cdot \left(\alpha \cdot E_{\tilde{t}}^{\frac{1}{\sigma}} \cdot \boldsymbol{\theta}_{\tilde{t}} - (-1+\alpha) \cdot E_{\tilde{t}} \cdot \boldsymbol{\theta}_{\tilde{t}}^{\frac{1}{\sigma}}\right)^2} \quad (4)$$

Assume that

$$a = (-1+\alpha) \cdot \alpha \cdot E_{\tilde{t}}^{-1+\frac{1}{\sigma}} \cdot \boldsymbol{\theta}_{\tilde{t}}^{-1+\frac{1}{\sigma}}$$

$$b = \left((1-\alpha) \cdot E_{\tilde{t}}^{\frac{-1+\sigma}{\sigma}} + \alpha \cdot \boldsymbol{\theta}_{\tilde{t}}^{\frac{-1+\sigma}{\sigma}}\right)^{\frac{2\sigma}{-1+\sigma}}$$

$$c = \left(\frac{\left((1-\alpha) \cdot E_{\tilde{t}}^{\frac{-1+\sigma}{\sigma}} + \alpha \cdot \boldsymbol{\theta}_{\tilde{t}}^{\frac{-1+\sigma}{\sigma}}\right)^{\frac{\sigma}{-1+\sigma}}}{\varphi_{\tilde{t}}}\right)^{2/\epsilon}$$

$$d = \epsilon \cdot \sigma \cdot \left(\alpha \cdot E_{\tilde{t}}^{\frac{1}{\sigma}} \cdot \boldsymbol{\theta}_{\tilde{t}} - (-1+\alpha) \cdot E_{\tilde{t}} \cdot \boldsymbol{\theta}_{\tilde{t}}^{\frac{1}{\sigma}}\right)^2$$

Hence, (4) can be given as $(abc)/d$. Upon closer inspection, we can see that for the assumed parameter values given in Table 1, $a$ and $d$ are always negative while $b$ and $c$ are always positive. Hence, $(abc)/d$ is always positive, i.e., the second principal minor is always positive.

From above, $|A_1| \leq 0, |A_2| \geq 0$ and $|A_k| = 0, \forall k > 2$. Hence, from proposition 1, the Hessian matrix $H_f(E_{\tilde{t}}^*, \theta_{\tilde{t}}^*, CAP_{i,v}^*, ACT_{i,v,\tilde{t}}^*)$ is negative semidefinite for all $E_{\tilde{t}}^*, \theta_{\tilde{t}}^*, CAP_{i,v}^*, ACT_{i,v,\tilde{t}}^* \in S_2$ where, $S_2$ is a feasible domain defined by constraints (1a-c). Therefore, from proposition 2, the function is concave.

If we add more electricity generation technologies, time periods and vintages then the resulting function will still be concave since from proposition 3, the sum of concave function will be concave.

**Proposition 7:** Feasible region defined by constraints (1a-c) is convex

Constraints (1a-c) are of the form $AX \geq b$. Since, all the constraints are linear, from proposition 4, the feasible domain is convex.

**Theorem 1:** The solution obtained by a nonlinear optimization solver is a global optimal solution.

The objective function $f(E_t, \theta_t, CAP_{i,v}, ACT_{i,v,t})$ is concave from proposition 6 and the feasible region is convex from proposition 7. Hence from proposition 5, the local maxima obtained from a nonlinear optimization solver (e.g. IPOPT from the Coin-OR initiative (Biegler and Zavala, 2009)) is a global maximum.

# Appendix B

**Data**

We modify an example energy system called 'utopia' to include solar and wind technology for electricity generation. This test case was, introduced in MARKAL, described in (Hewells et al., 2011) and extended in (Lavingne, 2017). Since the energy efficiency model is described for electric sector, we run the base case utopia model described by (Hewells et al., 2011) to determine the quantity demanded for electricity to meet the heating, lighting and transportation demand given in (Hewells et al., 2011).

In the modified utopia example, a single region is represented which has electricity demand. The electricity demand fluctuates depending on the season and time of the day: in general, more electricity is required during the day time and in winter. To generate electricity, six different power stations are available: coal (tech_COAL), nuclear(tech_NUC), hydro (tech_HY), diesel (tech_DSL), solar (tech_SOL) and wind (tech_WND). Diesel is imported (IMPDSL) and/or produced by a refinery (SRE) that converts imported crude oil (IMPOIL). Uranium and coal are also imported (via technologies IMPURN and IMPCOAL, respectively). The nuclear technology (tech_NUC) also take in fossil equivalent (FEQ) imported as (IMPFEQ) along with (URN) for electricity generation. The network diagram for the 'utopia' energy system is given in Appendix Figure 1.

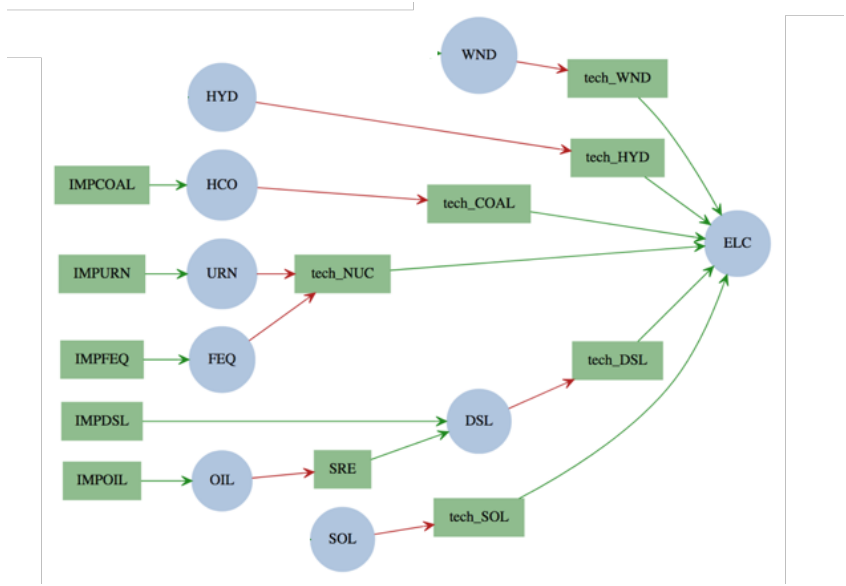

**Appendix Figure 1:** Graphical representation of a modified version of a test case called 'utopia' (developed for MARKAL). Energy technology is represented by green arrows, flow out by red arrows. Energy sources are shown on the left edge of the diagram (i.e., the import technologies), and on the right edge are the end-use electricity demand. This image was dynamically generated with an open source graphing utility called Graphviz.

The basic data used to calibrate the utopia application are summarized in Appendix Table 1. The future time horizon is from 1990 to 2010 while vintages of existing capacity is from 1960 to 1980. We assume that capacity factor of solar is zero during night time. Cost and performance data for wind and solar is taken from (EIA, 2018)

**Appendix Table 1:** Technology specifications for utopia database

| Parameter | Input | Efficiency | Output | Capital Cost | Variable cost | Fixed cost | Capacity to activity | Capacity factor | Life | Existing capacity | | |
|---|---|---|---|---|---|---|---|---|---|---|---|---|
| | | | | | | | | | | 1960 | 1970 | 1980 |
| Unit | | | | $M/GW | $M/PJ | $M/GW | | | years | GW | GW | GW |
| Technology | | | | | | | | | | | | |
| Tech_COAL | Coal | 0.32 | Electricity | 2000 | 0.3 | 40 | 31.54 | 0.8 | 40 | 0.175 | 0.175 | 0.15 |
| Tech_NUC | Uranium | 0.4 | Electricity | 4000 | 1.5 | 500 | 31.54 | 0.8 | 40 | 0 | 0 | 0 |
| Tech_NUC | FEQ | 0.32 | Electricity | 4000 | 0 | 0 | 31.54 | 1 | 1000 | 0 | 0 | 0 |
| Tech_HYD | Hydro | 0.32 | Electricity | 3000 | 0 | 75 | 31.54 | 0.275 | 100 | 0 | 0 | 0.1 |
| Tech_DSL | Diesel | 0.294 | Electricity | 1000 | 0.4 | 30 | 31.54 | 0.8 | 40 | 0.005 | 0.005 | 0.2 |
| Tech_WND | Wind | 0.34 | Electricity | 1600 | 12 | 41 | 31.54 | 0.4 | 40 | 0 | 0 | 0 |
| Tech_SOL | Solar | 0.34 | Electricity | 2000 | 11 | 25 | 31.54 | 0.4 | 40 | 0 | 0 | 0 |
| **Unit** | | | | $M/PJ/a | $M/PJ/a | $M/PJ/a | | | years | PJ/a | PJ/a | PJ/a |
| IMPDSL | | 1 | Diesel | 0 | 5 | 0 | 1 | 1 | 1000 | N/A | N/A | N/A |
| IMPCOAL | | 1 | Coal | 0 | 2 | 0 | 1 | 1 | 1000 | N/A | N/A | N/A |
| IMPOIL | | 1 | Oil | 0 | 8 | 0 | 1 | 1 | 1000 | N/A | N/A | N/A |
| IMPURN | | 1 | Uranium | 0 | 2 | 0 | 1 | 1 | 1000 | N/A | N/A | N/A |
| SRE | Oil | 1 | Diesel | 100 | 10 | 0 | 1 | 1 | 50 | N/A | N/A | N/A |

| **Year** | 1990 | 2000 | 2010 | For season: | Inter | | Summer | | Winter | |
|---|---|---|---|---|---|---|---|---|---|---|
| **Demand** | PJ/a | PJ/a | PJ/a | For time slice: | Day | Night | Day | Night | Day | Night |
| Electricity | 6.42 | 29.85 | 35.17 | | 0.1256 | .0594 | 0.0755 | 0.0344 | 0.4801 | 0.2248 |
| | | | | Year split: | 0.1667 | 0.0833 | 0.1667 | 0.0833 | 0.3333 | 0.1667 |

**Policy Analysis**

Welfare gain from efficiency credit scenarios as compared to welfare gain from 40 $/ton carbon tax scenario is shown in Appendix Figure 2. The results show that 42% of the welfare can be recovered with 10% efficiency credit as compared to the welfare gain from carbon tax of 40 $/ton.

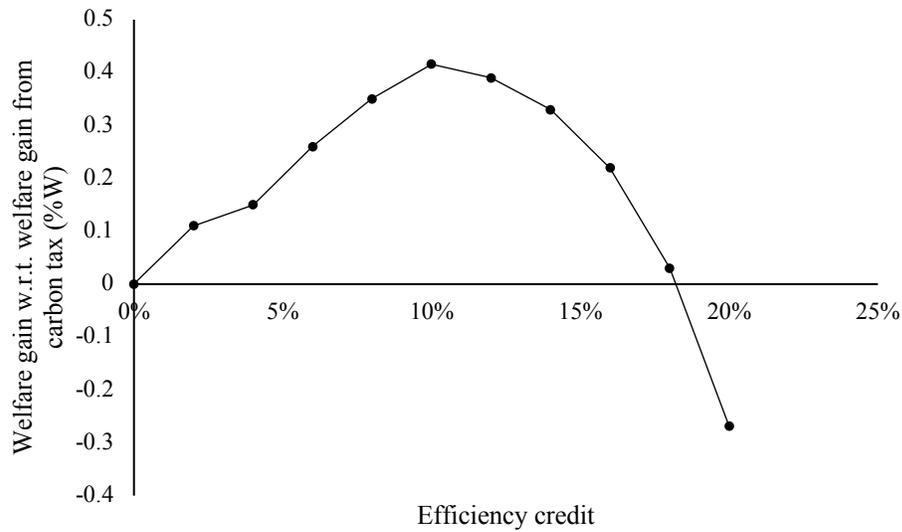

**Appendix Figure 2:** Welfare gain with respect to welfare gain with carbon tax = 40 $/tonCO2

Capacity expansion for different policy scenarios is shown in Appendix Figure 3. Majority of the capacity expansion is in coal power plant since it is the cheapest available technology. Capacity expansion of Hydro and crude oil processor (SRE) is due to the lower limit set on the capacity expansion in the utopia database. As can be observed, capacity expansion is lowest in carbon tax scenario. Also, increase in efficiency credit leads to lower electricity demand. Hence, as a result, overall capacity expansion decreases with increase in efficiency credit.

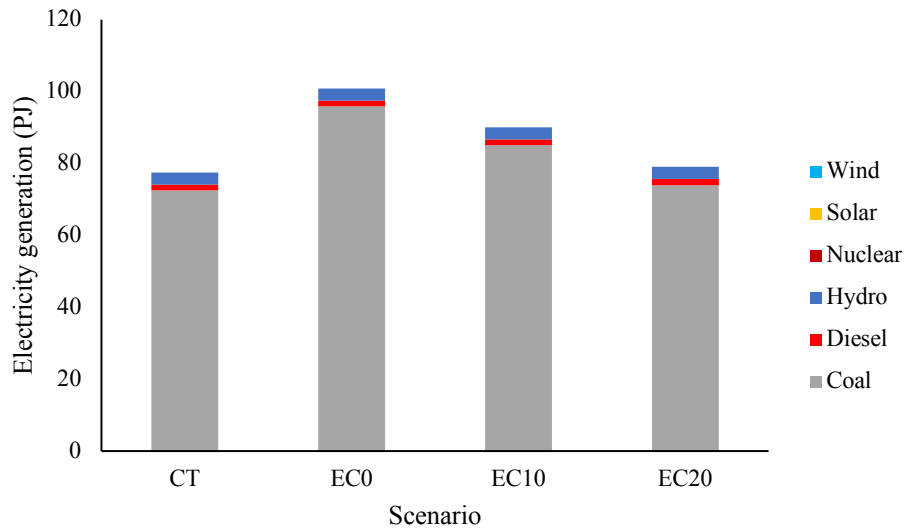

**Appendix Figure 3:** Electricity generation (PJ) for 40 $/ton Carbon tax (CT), no efficiency credit (EC0), 10% efficiency credit (EC10) and 20% efficiency credit (EC20)

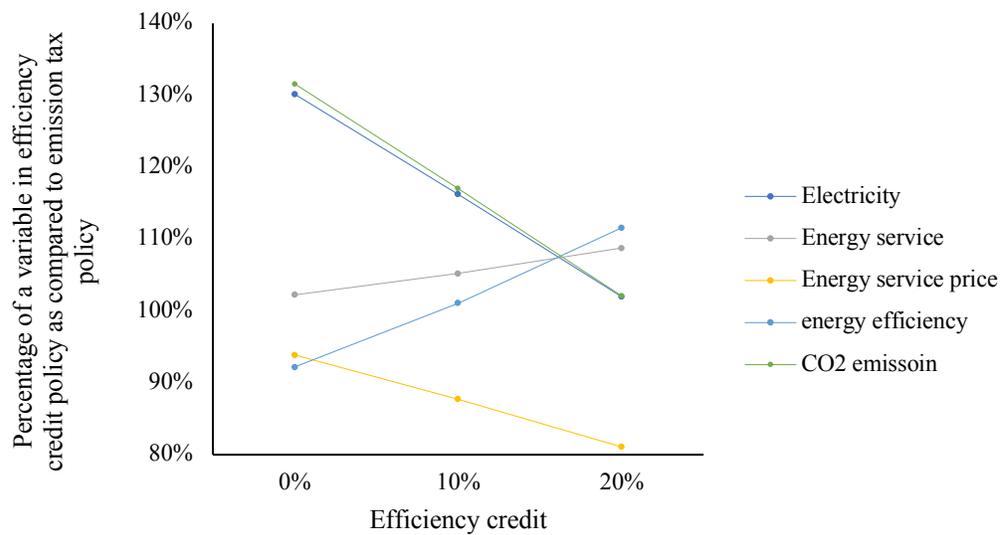

**Appendix Figure 4:** For a given efficiency credit a point of a variable represents the percentage quantity of a variable as compared to the quantity of a variable under emission tax policy